\documentclass[iop,apj,appendixfloats]{emulateapj}
\usepackage{graphicx}
\usepackage{amsmath}
\usepackage{amssymb}
\usepackage{color}

\gdef\xx[#1]{\textcolor{red}{#1}}

\gdef\kms{km\,s$^{-1}$}
\gdef\msun{M$_{\odot}$}

\gdef\blob{NGC1052--DF2}
\gdef\natpap{vD18a}
\gdef\gcpap{vD18b}

\newcommand{\GG}[1]{}

\lefthead{van Dokkum et al.}
\righthead{}
\slugcomment{ApJ Letters, in press}
\begin{document}


\title{The Distance of the Dark Matter Deficient Galaxy \blob}

\author{Pieter van Dokkum\altaffilmark{1},
Shany Danieli\altaffilmark{1},
Yotam Cohen\altaffilmark{1},
Aaron J.\ Romanowsky\altaffilmark{2,3},
Charlie Conroy\altaffilmark{4}
\vspace{8pt}}

\altaffiltext{1}
{Astronomy Department, Yale University, 52 Hillhouse Ave,
New Haven, CT 06511, USA}
\altaffiltext{2}
{University of California Observatories, 1156 High Street, Santa
Cruz, CA 95064, USA}
\altaffiltext{3}
{Department of Physics and Astronomy, San Jos\'e State University,
San Jose, CA 95192, USA}
\altaffiltext{4}
{Harvard-Smithsonian Center for Astrophysics, 60 Garden Street,
Cambridge, MA, USA}

\begin{abstract}

We recently inferred that the galaxy \blob\ has 
little or no dark matter and a rich system of unusual globular clusters.
We assumed that the galaxy is
a satellite of the luminous elliptical galaxy NGC1052 at $\approx 20$\,Mpc,
on the basis of its surface brightness fluctuations (SBF) distance of $19.0\pm 1.7$\,Mpc,
its radial velocity of $\approx 1800$\,km/s, and its projected position.
Here we analyze the color-magnitude diagram (CMD) of \blob, following
the suggestion by Trujillo et al.\ (2018) that the
tip of the red giant branch
(TRGB) can be detected in currently available HST
data and the galaxy is at $\sim 13$\,Mpc.
Using fully populated galaxy models 
we show that the CMD is strongly influenced by blends.
These blends produce a ``phantom'' TRGB
$\sim 2$ times brighter than the true TRGB, which can lead to erroneous distance
estimates $\sim 1.4$ times smaller than the actual distance.
We compare \blob\ to model images as well as other galaxies in our HST sample, and show that 
the large population of unblended RGB stars expected for distances
of $\sim 13$\,Mpc is not detected.
We also provide a new distance measurement to \blob\ that is free of calibration uncertainties, by anchoring
it to a satellite of the megamaser host galaxy NGC4258. From a megamaser-TRGB-SBF distance
ladder we obtain
$D=18.7\pm 1.7$\,Mpc, consistent with our previous measurement and with the distance to the
elliptical galaxy NGC1052.

\end{abstract}


\section{Introduction}

\hyphenation{kruijssen}


We recently identified a galaxy
with little or no dark matter ({van Dokkum} {et~al.} 2018a, hereafter \natpap).
\blob, originally discovered by {Fosbury} {et~al.} (1978), is a quiescent, spheroidal
``ultra diffuse'' galaxy (UDG; {van Dokkum} {et~al.} 2015)
with an effective radius of $R_e=2.2$\,kpc, a central
surface brightness $\mu(V_{606},0)=24.4$\,mag\,arcsec$^{-2}$, and
a stellar mass of $M_{\rm stars}\approx{}2\times{}10^8$\,\msun.
It has a remarkable population of globular clusters that rival
$\omega$\,Centauri in their luminosities, sizes, and ellipticities
({van Dokkum} {et~al.} 2018b, hereafter \gcpap). The globular cluster system
has an average radial velocity of
$\langle{}v\rangle\approx 1800$\,\kms\ and a velocity dispersion of
$\sigma_{\rm intr}=5.6^{+5.2}_{-3.8}$\,km/s (see {van Dokkum} {et~al.} 2018c).
This dispersion is similar to that expected from the stellar mass alone ($\sigma_{\rm stars}
=7.0^{+1.6}_{-1.3}$\,km/s), and using generative Jeans modeling in a Bayesian framework
{Wasserman} {et~al.} (2018) derive a 90\,\% upper limit of $M_{\rm halo}<1.2\times 10^8$\,\msun,
for a wide prior on the halo mass. Martin et al.\ (2018) find similar values for
the velocity dispersion (somewhat depending
on the assumptions), although they argue for weaker constraints on the total amount of
dark matter that could be present.

Most of these aspects depend on the distance that is assumed for the galaxy. There
is circumstantial evidence for a distance of $\approx 20$\,Mpc: it is located only
$14\arcmin$ away from the luminous elliptical galaxy NGC1052, which has distance
measurements ranging from 19.4\,Mpc to 21.4\,Mpc ({Tonry} {et~al.} 2001; {Blakeslee} {et~al.} 2002),
and its radial velocity implies a distance of $25\pm 1$\,Mpc if it is at rest
with respect to the Hubble flow. However, as noted in \natpap,
the properties of the galaxy are less extreme if it is closer to us. In particular,
the peak of the contamination-corrected
globular cluster luminosity function would
coincide with the canonical value for a
distance of $\approx 10$\,Mpc. The ratio of dark matter to luminous matter
would also be closer to expectations (although still low).
A distance of $\sim 10$\,Mpc would imply a 
peculiar velocity of order $\sim 1000$\,km/s for the galaxy, but it is difficult
to argue that this is less likely than having a population of extreme globular clusters
and an unusually low dark matter content.

\begin{figure*}[htbp]
  \begin{center}
  \includegraphics[width=0.95\linewidth]{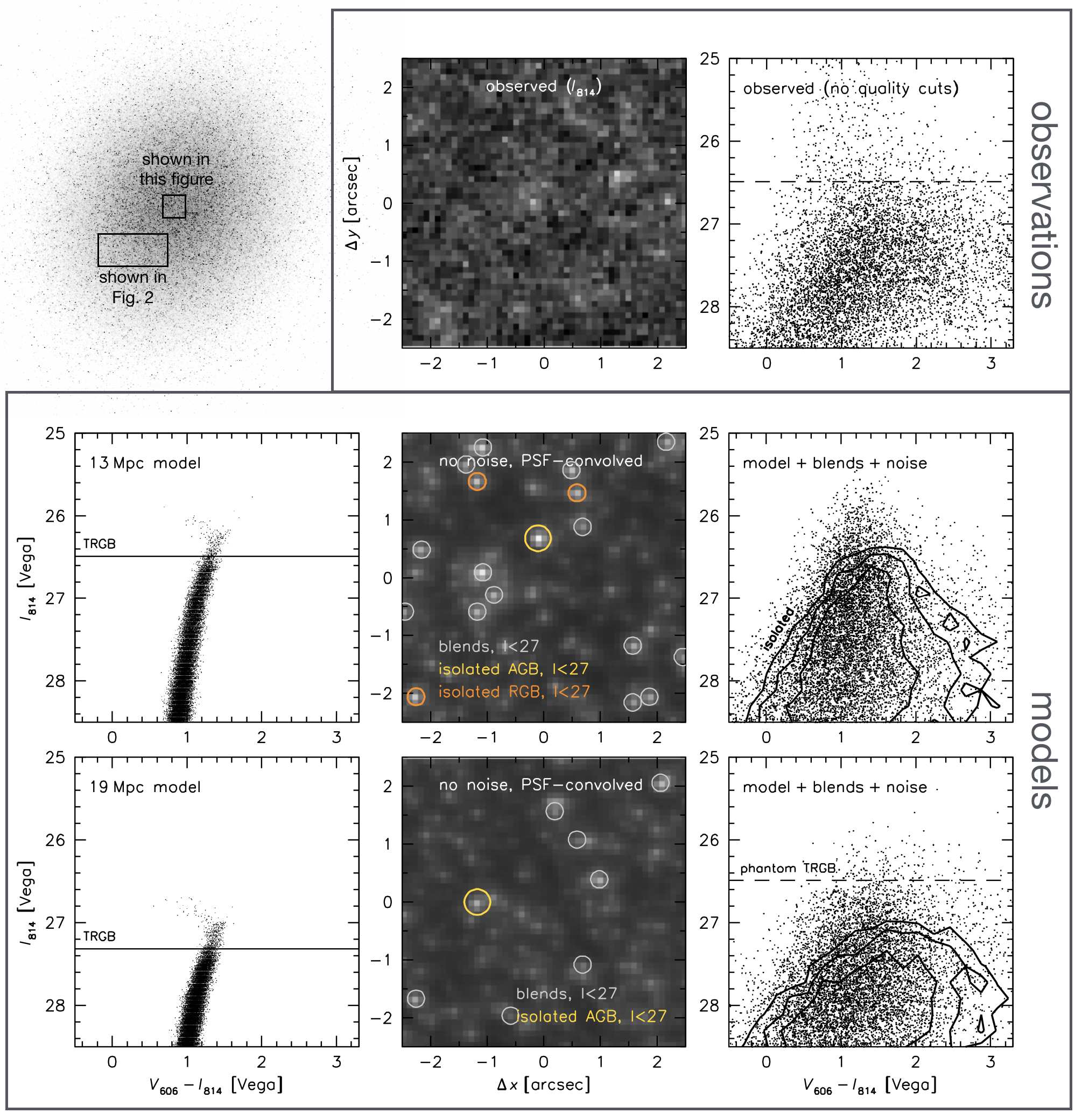}
  \end{center}
\vspace{-0.2cm}
    \caption{
Comparison of the CMD of \blob\ (top right; prior to applying quality cuts to eliminate blends)
to models at distances of 13\,Mpc (middle row)
and 19\,Mpc (bottom row). The models reproduce the global properties of \blob.
The observed ``raw'' CMD shows a ridge line at $I_{814}\approx 26.5$.
The images in the middle column show the central $5\arcsec
\times 5\arcsec$ of the data and the models. Circles indicate detectable isolated
stars and blends:
most of the brightest compact objects are blends.
The distributions
of isolated stars in the CMDs
are indicated with contours in the panels on the right.
Taking blends
and photometric errors into account the model CMD for 19\,Mpc is a good match
to the data, including the ``phantom'' TRGB at $I_{814}\approx 26.5$.
}
\label{theory.fig}
\end{figure*}

In \natpap\ we argued that we do not detect individual red
giant branch (RGB) stars in \blob, and attributed this to the
large distance of the galaxy.
For  distances $\gtrsim 15$\,Mpc, individual
giants are undetected in single-orbit HST images
but blend into
surface brightness fluctuations (SBF) where the stellar density is high enough.
Hence we used SBF in the inner parts of the galaxy
to determine the distance to \blob, arriving at $19.0\pm 1.7$\,Mpc.
{Trujillo} {et~al.} (2018) (hereafter T18) suggest
that individual
RGB stars {\em are} detected in the
HST imaging of \blob. They detect many
compact objects, and identify a sharp increase in the number
of detections below $I_{814}\approx 26.5$. Interpreting this ridge in the CMD as
the TRGB, the distance they find is $13.1\pm{}0.8$\,Mpc.
They also cast doubt on the SBF distance that was derived in vD18,
suggesting that calibration errors led to an overestimate of the distance.

In this {\em Letter} we analyze the CMD of \blob\ and 
show that blends produce a ``phantom'' TRGB that is brighter than the true TRGB.
We also derive a distance to \blob\ that is independent of the absolute
calibration of the SBF signal. This paper is a companion to a  study of
all 23 galaxies in our Cycle 24 HST program ({Cohen} {et~al.} 2018, hereafter
C18).

\begin{figure*}[htbp]
  \begin{center}
  \includegraphics[width=0.9\linewidth]{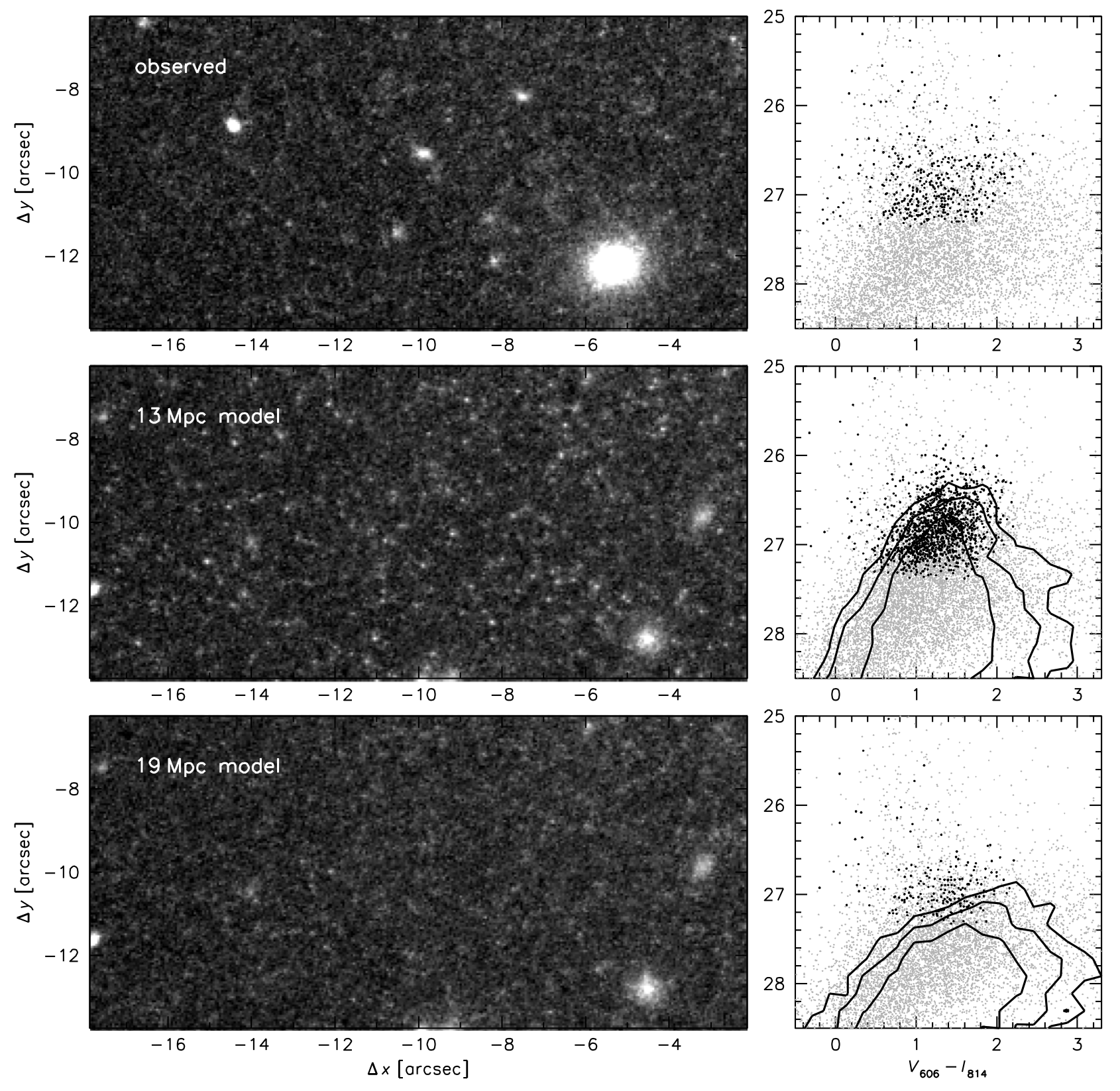}
  \end{center}
\vspace{-0.2cm}
    \caption{
Comparison of \blob\ to artificial galaxy images placed in the HST/ACS
data  (see Fig.\ 1
for their location). The CMDs measured with DOLPHOT are shown at right.
Grey points are ``raw'' photometry; black points are objects that survive
standard  quality cuts. Contours are repeated from Fig.\ 1 and show the
expected distribution of isolated stars.
For a distance of 13\,Mpc the HST images would have
shown a large number of isolated stars above the deteciton limit.
}
\label{empirical.fig}
\end{figure*}

\section{Modeling the Color-Magnitude Distribution}

\subsection{Observed CMD}

The ``raw'' distribution of detected sources in the
CMD of \blob\ is shown in the top right panel of Fig.\ \ref{theory.fig}.
The photometric analysis was done using the ACS module of DOLPHOT, which itself
is based
on HSTPHOT ({Dolphin} 2000). DOLPHOT operates directly on the
{\tt flc} files.
Our methodology is outlined in {Danieli} {et~al.} (2017) and C18; we follow
identical procedures to those established in {Dalcanton} {et~al.} (2009) for crowded ACS
photometry, as described in detail in the DOLPHOT
manual.\footnote{http://americano.dolphinsim.com/dolphot/dolphotACS.pdf}

DOLPHOT measures various parameters of the detected sources (such as their sharpness and
degree of crowding) in order to remove spurious detections and blends
(see, e.g., Fig.\ 2 in {M{\"u}ller}, {Rejkuba}, \&  {Jerjen} 2018). The ``raw'' CMD
shown in Fig.\ \ref{theory.fig}
includes all DOLPHOT detections within a radius of $R=2R_e$,
before applying any of these quality cuts. DOLPHOT detects
many sources in \blob: 1609 with $I_{814}<27$. Furthermore, there is a conspicuous
ridge in the CMD at $I_{814}\approx 26.5$, indicated with the dashed line.
Interpreting this ridge as the TRGB would imply a distance of $\sim 13$\,Mpc (T18, C18).
but we demonstrate below that it is spurious.

\begin{figure*}[p]
  \begin{center}
  \includegraphics[width=0.95\linewidth]{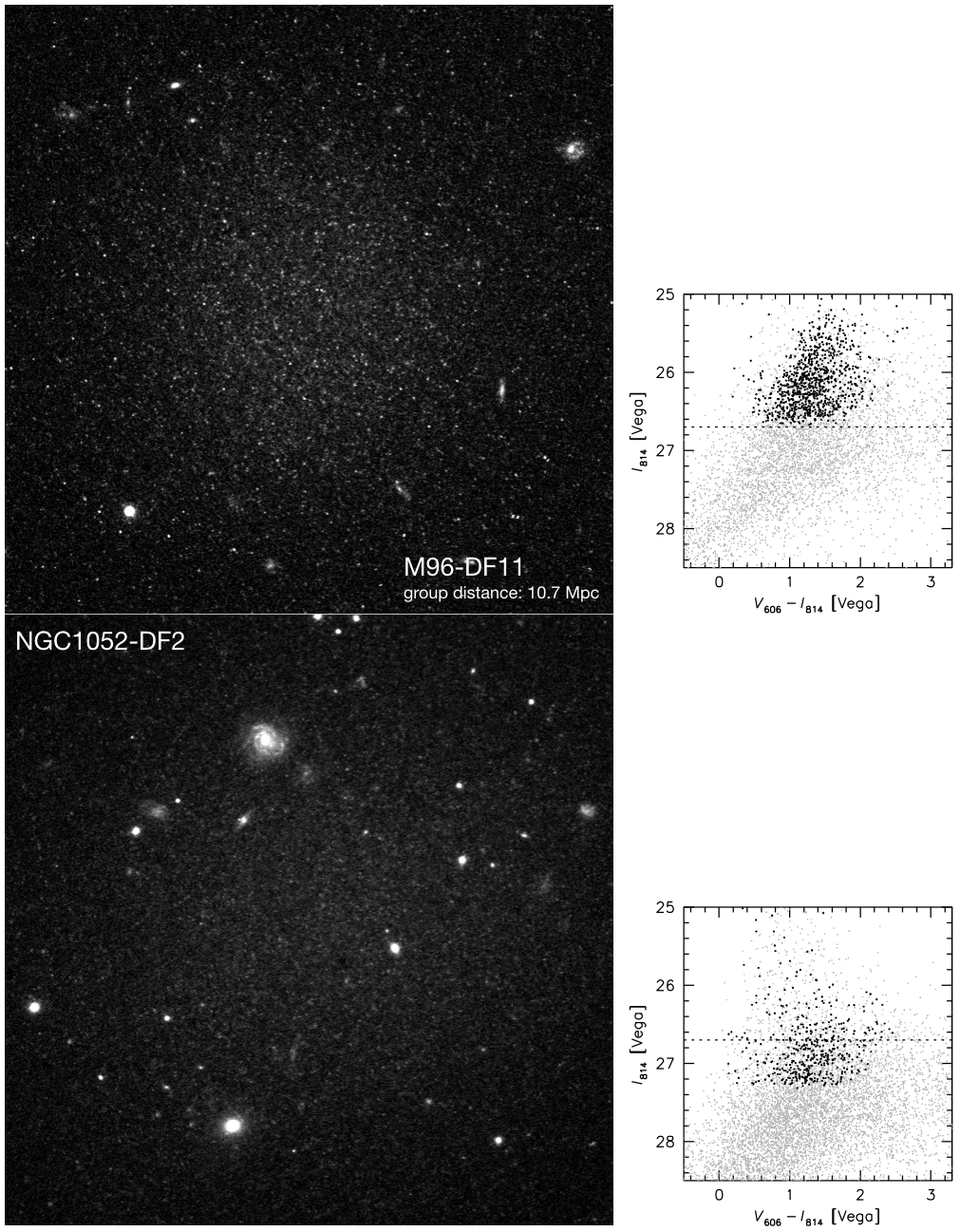}
  \end{center}
\vspace{-0.2cm}
    \caption{
\blob\ and M96-DF11 (C18)
have very similar surface brightness, size, morphology, and
integrated color. The images span $21\arcsec\times{}21\arcsec$. Grey dots in the CMD
are all detections; black dots are what remains after quality cuts. The broken line
indicates the depth of the (half-orbit) M96-DF11 $I_{814}$ data.
For M96-DF11 we reach below the tip of the giant branch, and the galaxy
takes on a resolved appearance. The distance to the M96 group is $10.7\pm{}0.3$\,Mpc
(see \S\,4). The distance to \blob\ is clearly much greater.
}
\label{df11_compare.fig}
\end{figure*}

\subsection{Modeled CMDs at 13\,Mpc and 19\,Mpc}

To understand the distribution of detections in the observed CMD
we generated
fully populated model galaxies with {\tt ArtPop}.
This code is described in detail in \S\,2 of {Danieli}, {van Dokkum}, \&  {Conroy} (2018). Briefly,
{\tt ArtPop} draws stars from the MIST isochrones ({Dotter} 2016; {Choi} {et~al.} 2016) for
a specified IMF and set of stellar population parameters, determines
the brightness of the stars in particular filters for a chosen distance,
and places them in an image according to a specified spatial distribution
(parameterized by the position, effective radius,
{Sersic} (1968) index, ellipticity, and position angle). The images are optionally
convolved with an instrumental PSF.
The \blob\ models are constrained to reproduce its observed integrated color and
2D surface brightness distribution, while varying the distance. Specifically,
the 19\,Mpc model has an age of 10\,Gyr, [Fe/H]$=-1$, and total magnitudes of
$V_{606}{\rm [AB]}=16.26$, $I_{814}{\rm [AB]}=15.84$. Its stellar mass is
$2.2\times 10^8$\,M$_{\odot}$ and the simulated image contains $10^9$ stars down
to $m_{814}=42.7$. The 13\,Mpc model is very similar but has
a stellar mass of $1.06\times{}10^8$\,M$_{\odot}$.
The CMDs of models at 13\,Mpc and at 19\,Mpc are shown in
the left panels of Fig.\ \ref{theory.fig}. The photometry was perturbed slightly
to limit overlap between the plotted points. For these distances the TRGB
is at $I_{814}=26.5$ and $I_{814}=27.3$ respectively, according
to the color-dependent
calibration of {Rizzi} {et~al.} (2007).

The central regions of the
{\tt ArtPop} images, convolved with the $I_{814}$ PSF,
are shown in the middle column.
Many compact objects are visible in the central
$5\arcsec\times 5\arcsec$ and throughout the images. {\em However, most of these
are not isolated stars but blends.} We identified blends by calculating the
flux contribution by
other stars with $I_{814}<29$ within a radius of $0\farcs 15$. If this contribution
exceeds 20\,\% the primary star and the contaminating stars are flagged as blends.
The circles in the simulated images indicate unblended stars with $I_{814}<27$,
below (orange) and above (yellow) the TRGB. There are only four unblended stars
in the central regions of the 13\,Mpc model, and only one in the 19\,Mpc model.
This result is not sensitive to the precise definition of blends: in both
models blends vastly outnumber isolated stars among the bright detections.

Next, we simulate the observed CMD by summing the fluxes of blended stars and adding
photometric noise (determined
using DOLPHOT's artificial star photometry; see C18). The results
are shown in the right panels of Fig.\ \ref{theory.fig}. The distribution of sources
in the CMD shows marked differences between the 13\,Mpc and 19\,Mpc models: at 13\,Mpc
the basic outline of the giant branch is conserved as a fairly narrow, near-vertical
plume of points, whereas at 19\,Mpc the distribution is broad and red. The 19\,Mpc
model reproduces the qualitative features of the observed distribution in the CMD,
including the ridge at $I_{814}\approx 26.5$. Stars near the ridge line are almost
exclusively blends of stars just below the true TRGB,
producing a mean boost to the flux of 0.6\,mag. This is not a new result: it is well
known that blends produce a ``phantom'' TRGB above the true tip, with the
distance between the true and phantom TRGB a function of the stellar density
(see, e.g., Fig.\ 4 in {Bailin} {et~al.} 2011). 
Very approximately, the boost is a factor of $\approx 2$, leading to a factor
of $\approx 1.4$ error in the distance.

\subsection{Analysis of Simulated Data}

As noted above, in nearly all studies of the CMD
detected objects are subjected to stringent quality cuts, in order
to mitigate the effects of blends and spurious sources (see, e.g., {Dalcanton} {et~al.} 2009; {Radburn-Smith} {et~al.} 2011; {McQuinn} {et~al.} 2017). 
We simulate these cuts, as well as the effects of background galaxies, image defects,
and non-linear noise, by placing the {\tt ArtPop} models in the
\blob\ ACS images and analyzing them in the same way as the actual data. The results
are shown in Fig.\ \ref{empirical.fig}.
The qualitative difference between the 13\,Mpc
model and the 19\,Mpc model is striking; as the data reach just below the TRGB for
13\,Mpc the simulated image shows many individual stars, whereas they remain undetected
for a distance of 19\,Mpc. The comparison to the data is unequivocal: the 13\,Mpc
model can be ruled out. The CMDs demonstrate this same result.
Using standard 
cuts\footnote{We use the crowding and sharpness cuts of {Radburn-Smith} {et~al.} (2011) 
and the signal-to-noise ratio cuts of {McQuinn} {et~al.} (2017).} 
 the vast majority
of detected objects disappear in both the \blob\ CMD 
and in the 19\,Mpc model  (1304 out of 1609 with $I_{814}<27$). Furthermore,
the distribution of
remaining sources in the 13\,Mpc model
closely follows the expected distribution of isolated stars.

\section{Comparison to M96-DF11}

The analysis in \S\,2 uses models to interpret the data. Owing to our relatively
large sample of low surface brightness objects (described
in C18) we can also perform direct
comparisons between HST images of similar-looking galaxies at different distances.
In particular, we obtained single-orbit (split between $V_{606}$ and $I_{814}$)
observations of
11 galaxies in the rich M96 group at 10.7\,Mpc
(see {Tully}, {Courtois}, \& {Sorce} 2016, and below).\footnote{The M96 (Leo) group has
an estimated spatial extent of 0.2\,Mpc (Tully 2015), or 2\,\% of the distance, which
means we can safely assume that all its members are at the same distance.}
The appearance of these galaxies is qualitatively different from that of \blob: they
resolve into a myriad of well-detected RGB stars.
We highlight M96-DF11 in Fig.\ \ref{df11_compare.fig}, as this galaxy has
very similar observed global properties as \blob: $R_e=16\arcsec$
($21\arcsec$ for \blob), $n=0.7$ ($0.6$), $\mu_{0,V}=24.0$ ($24.2$), $b/a=0.95$ ($0.85$),
and $V_{606}-I_{814}=0.45$ ($0.40$).
The resolved appearance of M96-DF11
and its CMD are dramatically different
from  \blob. Specifically, the number of detected stars with $I_{814}<26.5$ is
$9\times$ higher (785 vs.\ 80).
For a distance of 13\,Mpc the equivalent limit is
$I_{814}=26.9$; we find 247 sources in \blob\ to that limit.
Figures \ref{empirical.fig} and \ref{df11_compare.fig} demonstrate that
we do not detect individual stars below the TRGB in \blob, ruling out distances as
close as $\sim 13$\,Mpc.

\begin{figure*}[htbp]
  \begin{center}
  \includegraphics[width=0.85\linewidth]{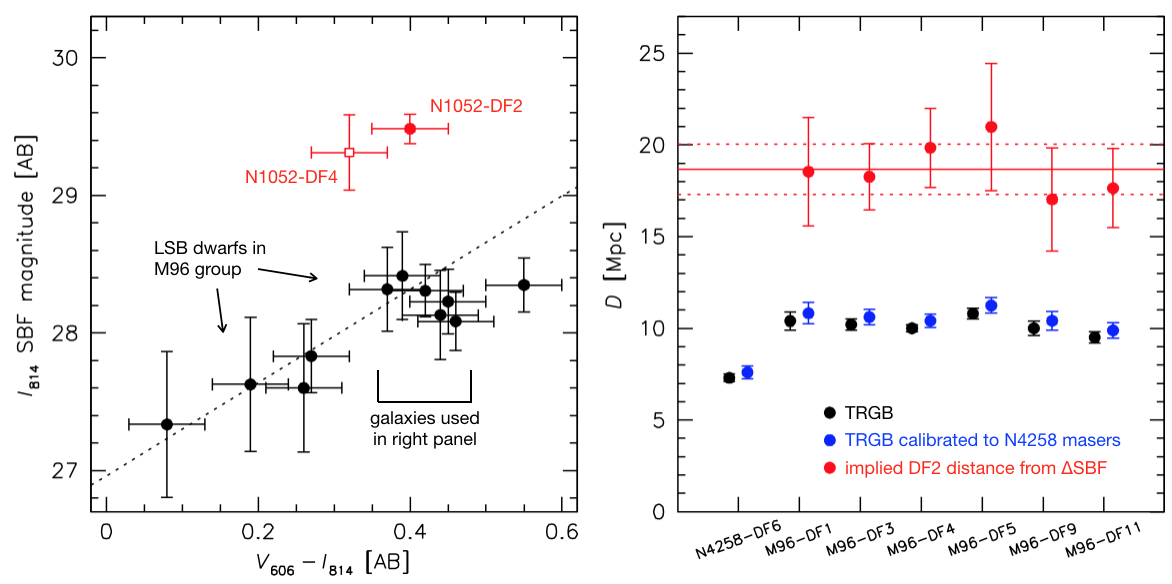}
  \end{center}
\vspace{-0.2cm}
    \caption{
Left: comparison of observed SBF magnitudes for \blob, NGC1052-DF4, and 11 low
luminosity galaxies in the M96 group.
The six galaxies with very similar colors as \blob\ are used to
determine $\Delta{\rm SBF}$ and hence
the relative distance between these galaxies and \blob.
The broken line shows the
extrapolation of the {Blakeslee} {et~al.} (2010) relation for $D=10.7$\,Mpc,
for reference only.
Right: TRGB distances to NGC4258-DF6 and six M96 galaxies with similar colors
as \blob\ (black points). Blue points are scaled
under the assumption that NGC4258-DF6 is at the megamaser distance of NGC4258.
Red points are measurements of the absolute distance to \blob\ based on the
blue points and the $\Delta{\rm SBF}$ values derived in the left panel.
}
\label{distance.fig}
\end{figure*}

\begin{figure*}[htbp]
  \begin{center}
  \includegraphics[width=0.85\linewidth]{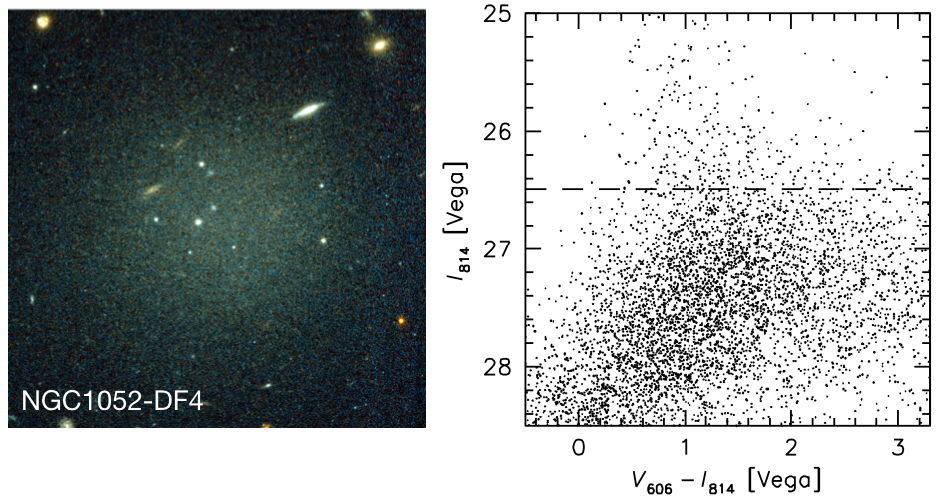}
  \end{center}
\vspace{-0.2cm}
    \caption{
Left: color image generated from the $V_{606}$ and $I_{814}$ data of
NGC1052-DF4, a galaxy with similar observed properties as \blob.
Right: the ``raw'' CMD is nearly identical to that of \blob\ (shown
in Fig.\ 1), as is its SBF magnitude. NGC1052-DF4 is the brightest
example of several galaxies that are at the same distance as
\blob\ and located (in projection) near the center of the NGC1052 group.
It is unlikely that they are all in a foreground structure.
}
\label{df4_comp.fig}
\end{figure*}

\section{A Distance for \blob\ Calibrated to H$_2$O Megamasers}

We have shown that the distribution of sources
in the CMD is qualitatively consistent with the \natpap\ SBF value of $19.0\pm 1.7$\,Mpc,
but it is hazardous to measure a quantitative distance from the blends and AGB stars
that constitute the detections. 
Our SBF measurement has a small random
uncertainty (see Fig.\ 8 in C18); however,
details in the methodology can lead to systematic errors ({Mei} {et~al.} 2005a;
Blakeslee \& Cantiello 2018),
particularly for low surface brightness galaxies ({Mieske}, {Hilker}, \& {Infante} 2003).
Furthermore,
the calibration that we use is
an extrapolation of relations that were established for more metal rich galaxies
(see {Mei} {et~al.} 2005b; {Blakeslee} {et~al.} 2010).
In this Section we make use of our sample of HST-observed low luminosity galaxies (C18) to
derive a distance to \blob\ that is insensitive to the details of the measurement technique
and does not rely on absolute stellar population calibrations.

\subsection{Relative Distance Between \blob\ and Dwarfs in the M96 Group}

The galaxies in the M96 group play a pivotal role in our analysis, as
the TRGB is unambiguously detected
in the CMDs and the galaxies are sufficiently well-populated for an accurate
measurement of the apparent SBF magnitude. In the left panel of Fig.\ \ref{distance.fig}
we show the apparent SBF magnitude as a function of the integrated color for the
11 galaxies and for \blob, with measurements
taken from C18. As expected, there is a correlation, such that bluer galaxies have
brighter SBF magnitudes. Six galaxies (including M96-DF11) have colors
that are nearly identical to that of \blob.
For each of the six we can obtain a measurement of
the relative distance between \blob\ and the M96 group without relying on an absolute
calibration of the SBF magnitude. The average offset
is $\Delta {\rm SBF}=1.23$ mag, corresponding to a distance ratio of 1.76.

The broken line shows the extrapolation of the 
{Blakeslee} {et~al.} (2010) relation that was used in \natpap\ and in C18, for
the Cosmicflows distance of 10.7\,Mpc to the M96 group ({Tully} {et~al.} 2016). We do not
use this relation in the present study but we note that it provides
a satisfactory description of the data.

\subsection{Absolute Distance to Dwarfs in the M96 Group, and to \blob}

Next we determine the absolute distance to these six dwarfs.
The black points in the right panel of Fig.\ \ref{distance.fig} show their
TRGB distances, taken from C18.
The methodology that we use for the TRGB measurements is
detailed in C18. Briefly, we use
the logarithmic edge-detection of {M{\'e}ndez} {et~al.} (2002), and derive TRGB
distances  using the color-dependent
calibration of {Rizzi} {et~al.} (2007).

Although we verified that our methodology
produces similar results as other methods (e.g., {Makarov} {et~al.} 2006) when run
on the same data, we do not use the TRGB distances directly as
we cannot exclude the possibility that
residual blending or other systematic effects influence our measurements.
Instead, we make use of the fact that our sample includes 
a satellite galaxy of NGC4258, NGC4258-DF6. It is at a projected distance of 57\,kpc and
it has a well-determined TRGB distance of $7.3 \pm 0.2$\,Mpc (random error only; see Fig.\ 7 in C18).
NGC4258 has an exquisitely well-established {\em absolute} distance from
the Keplerian rotation of its H$_2$O megamasers:
$D=7.60\pm 0.23$\,Mpc  ({Humphreys} {et~al.} 2013). We therefore
apply a correction factor of $7.6/7.3=1.04$ to the TRGB distances
of the six M96 dwarfs to bring them on the megamaser system (blue
points in Fig.\ \ref{distance.fig}). The average maser-calibrated TRGB
distance to these galaxies is $10.7\pm 0.4$\,Mpc, in excellent agreement with the
canonical distance to the M96 group of $10.7\pm 0.3$\,Mpc ({Tully} {et~al.} 2016).

Adopting a common distance of $10.7\pm 0.4$ Mpc for the six galaxies,
we use the difference between their SBF magnitude and that of
\blob\ to obtain six absolute distance estimates of \blob\ (red points in
Fig.\ \ref{distance.fig}).  The average value, with propagated errors, is
$D=18.7 \pm 1.7$\,Mpc. The error includes a 3\,\% uncertainty due to the unknown
distance between NGC4258-DF6 and NGC4258 itself
(based on the extent of the M31 satellite system; see {Conn} {et~al.} 2012),
and an estimated 0.1\,mag scatter in the absolute SBF magnitude at fixed color (see {Blakeslee} {et~al.} 2010).
This distance is
insensitive to
methodological details, and entirely free from stellar
population-based absolute calibrations (although consistent with them).

\section{Discussion}

In this {\em Letter}
we demonstrated that individual stars fainter than the tip of the giant branch are
not detected in the current HST imaging of \blob\ and
determined an SBF distance to \blob\ of $18.7 \pm 1.6$\,Mpc that is independent of
calibration uncertainties.
We infer
that the galaxy is probably a satellite of the massive elliptical galaxy NGC1052
at $D=20.4\pm 1.0$\,Mpc ({Tonry} {et~al.} 2001; {Blakeslee} {et~al.} 2002).
This conclusion is supported by an independent analysis of the SBF signal by Blakeslee
\& Cantiello (2018), who find $D=20.4\pm 2.0$\,Mpc.

As noted in the Introduction and in T18 the unusual dark matter content and globular cluster
population are more easily explained with a smaller distance, but the luminosity
of the red giants ``overrules'' such indirect arguments.
T18 emphasize that a smaller distance resolves {\em both} the unusual luminosities and the unusual
sizes of the globular clusters in \blob, but we note here that these properties are
probably coupled, with the characteristic luminosity and size of clusters possibly
set in tandem by the large scale environment (Reina-Campos \& Kruijssen 2017).
Specifically, as discussed
in \gcpap, the same gas pressures were needed to form
the globular clusters in \blob\ and
those in the Milky Way
(e.g., Elmegreen \& Efremov 1997).


Finally, we emphasize
that \blob\ is not alone: it is one of several low surface
brightness galaxies in the NGC1052 field that are all consistent
with the same distance (see C18). They are all located near NGC1052 in the
Dragonfly frame (see Fig.\ 1 in C18)
and none of them have a detected red giant branch in our $I_{814}$ images.
In Fig.\ \ref{df4_comp.fig} we
show the brightest of these other galaxies, 
NGC1052-DF4. This galaxy has a similar
morphology, size, and surface brightness as \blob.
The SBF magnitudes (Fig.\ \ref{distance.fig}) and CMDs
(Figs.\ 1 and \ref{df4_comp.fig})
of the two galaxies
are nearly identical. It is unlikely that all
these galaxies are associated with the possible foreground spiral
NGC1042, while there would be
no galaxies of this apparent brightness
that are actually associated with the rich NGC1052 group.

We conclude that the red giant population of \blob, analyzed here through the CMD
and surface brightness fluctuations, implies a distance of $\sim 20$\,Mpc.
With significantly deeper HST data than are available now
it should be possible to measure the TRGB in the outskirts of the galaxy,
for a definitive distance with an accuracy of $\sim 5$\,\%.


\acknowledgements{
We thank the other members of Team Dragonfly for their help, and John Blakeslee and Oliver M\"uller
for their comments. Support from {\em HST} grant HST-GO-14644
and NSF grants AST-1312376, AST-1616710, 
AST-1518294, and AST-1613582
is gratefully acknowledged.
}


\end{document}